\documentclass[aip,apl,graphicx]{revtex4-1}
\usepackage{graphicx}
\usepackage{dcolumn} 
\usepackage{bm}
\usepackage{float}

\begin{document}
 
\title{Electron and thermal transport via Variable Range Hopping  in MoSe$_{2}$ single crystals}

\author{Dhavala Suri}

\author{R. S. Patel}
\email{rsp@goa.bits-pilani.ac.in}
\affiliation{Department of Physics, \\ Birla Institute of Technology \& Science Pilani – K K Birla Goa Campus, \\Zuarinagar, Goa – 403 726, India}

\begin{abstract}
Bulk single crystal Molybdenum diselenide has been studied for its electronic and thermal transport properties. We perform resistivity  measurements with current in-plane~(CIP) and current perpendicular to plane~(CPP) as a function of temperature. The CIP measurements exhibit  metal to semiconductor transition at  $\simeq 31$ K. In the semiconducting phase ($T > 31$ K), the transport is best explained by variable range hopping (VRH) model. Large magnitude of resistivity in CPP mode indicates strong structural anisotropy.  Seebeck coefficient as a function of temperature  measured in the range $90 - 300$ K, also agrees well with the VRH model.  The room temperature Seebeck coefficient is found to be $139$ $\mu$V/K.   VRH fittings of the resistivity and Seebeck coefficient data indicate high degree of localization.
 
\end{abstract}

\maketitle

\section{Introduction}

Transition metal dichalcogenides (TMDCs) have significantly contributed to the recent advances in two dimensional  materials. They are layered  materials composed of transition metal (M) and chalcogens (X) in the form MX$_2$. TMDCs provide a wide platform to explore highly interesting quantum phases of matter like charge density waves, superconducting phase transition, metal to insulator transitions and so on. They possess advantageous properties for nanodevice applications like large Seebeck coefficient, non-saturating magnetoresistance, superconductivity, tunable band-gap etc. \cite{Wang,mali,splend,qi,mak,radi,Ovchinnikov2014,Qiu2015}.  This makes TMDCs highly pertinent for research both fundamentally and technologically. In this work, we explore Molybdenum Diselenide (MoSe$_2$) in its intrinsic semiconducting phase. MoSe$_2$ has found  wide variety of applications in optoelectronic and photovoltaic nanodevices \cite{Tongay2012,Wang}.  Field effect transistors of MoSe$_2$ display a large on-off ratio ($\sim 10^6$) and significant mobility ($100-160$ cm$^{2}$ V$^{-1}$ s$^{-1}$)\cite{Larentis,Chamla,Pradhan}.  MoSe$_2$ has also been reported to have a negative magnetoresistance \cite{Dau}. It exhibits exotic phenomena that are rich in fundamental physics \citep{Kumar,Zhao,Tongay2012,Bernede,Wieting1980}. We conduct a study on bulk MoSe$_2$  single crystals and present intriguing features which are of high scientific interest.

We study electron and thermal transport properties of  MoSe$_2$ through resistivity and Seebeck coefficient measurements. Resistivity measurements as a function of temperature reveal scattering mechanism involved at various temperature regime. 
We infer that the appropriate transport mechanism that best explains the data is variable range hopping (VRH) mechanism proposed by Mott in 1969 \cite{Mott1969,Mottbook,Mottbook2,Mottrmp}.  According to this model, conduction of charge carriers occurs via hopping mechanism. The electrons hop between localized states in the crystal \cite{halperin}, for which the necessary energy  is provided by phonons. The resistivity is given by,
\begin{eqnarray}
\rho(T)=\rho_0 \ \exp\left( \frac{T_0}{T}\right)^\frac{1}{d+1}
\end{eqnarray}

 where $d$ is the dimensionality. In this model  density of states close to the fermi level is assumed to be constant. Using the VRH model, Seebeck coefficient calculations by Chaikin et.al.  \cite{chaikin}  lead to a temperature dependence of Seebeck coefficient given by, 
 \begin{eqnarray}
 S(T)=\frac{k_B}{2e}\left[\frac{\Gamma(d/2 + 1)}{D_0(\mu)\pi^{d/2}}\right]^{2/(d+1)} \left(\frac{2\alpha}{d}\right)^{2d/(d+1)} \left[\frac{\partial ln D_0(\epsilon)}{\partial \epsilon}\right]_{\epsilon=\mu} \left(k_B T\right)^{(d-1)/(d+1)}
 \end{eqnarray} 
 where, $D_0(\mu)$ is the density of states,   $D_0(\epsilon)$ is the density of states when interaction is turned off, $\alpha$ is the decay length of the  wave function considered and $d$ is the dimension. Hence the overall dependence of $S$ on $T$ will  be given by,
 \begin{eqnarray} S(T) \propto T^{\frac{d-1}{d+1}} \end{eqnarray}

We present a detailed experimental study of resistivity measurements for current in-plane (CIP) mode and current perpendicular to plane (CPP) mode, the schematics of which are shown in Fig. 1(a). Resistivity in CIP mode shows a very intriguing feature of metal to semiconductor transition at low temperature. At higher temperatures the transport exhibits semiconducting phase and the mechanism is primarily through hopping between localized states. Measurements in CPP mode also show semiconducting phase with conduction via hopping. We perform in-plane Seebeck coefficient measurements (Fig. 1(b)) for which the data fits best for the VRH model. 

It has been reported that two dimensional nature of the material, supports larger Seebeck coefficient \cite{Imai}. Large magnitude of Seebeck coefficient encourages us to look for properties which might support Phonon Glass Electron Crystal (PGEC)  behavior in this material \cite{pgec1,Beekman,Wu2014,Snyder2008}.  This concept aims to engineer materials with    low thermal conductivity (high phonon-phonon scattering as in glass) and high electronic conductivity (perfect crystal like behavior for electrons). TMDCs might satisfy this requirement due to their anisotropic crystal structure \cite{Beekman}. In further sections we present the experimental details and the results of the study, and lastly conclusions.

\section{Experimental details}

We perform DC resistivity measurements in CIP and CPP modes  using standard four probe technique. Typical sample dimensions are:  thickness  $ 0.1$ mm and area $3$ mm $\times$ $2$ mm. The sample is mounted on the chip customized for measurements in closed cycle He cryostat (Oxford Cryomech PT405). Oxford MercuryiTC temperature controller is used along with the cryostat, for temperature dependent measurements between 9-300 K. Keithely 2401 sourcemeter and Keithley 2182A nanovoltmeter are used to source current and measure voltage respectively. Contacts are made using silver paste on the same side of the sample for CIP measurements and on opposite sides for CPP measurements as shown in fig. 1 (a). 

The Seebeck measurements are performed in an in-house built dip-stick which is calibrated using measurements on standard samples. As shown in fig. 2 (b) resistivity dip stick consists of a long stainless steel tube whose one end consists of the sample mounting pins. Wires drawn out from these pins are connected to  external hardware by  means of electrical feed through at the other end. The tube has an extension to vacuum pump. The sample end is sealed in a copper can and maintained in a vacuum of the order of 10$^{-3}$ mbar for temperature dependent measurements. The sample mount is shown in detail in fig. 2 (c).  The sample is placed between two thermal blocks separated by an air gap of $1$ mm (Fig. 1 (c)). Copper wires are used as probes to measure thermal voltage. Contacts are made on the sample using silver paste. Miniature temperature sensors (Pt-100) are placed in close proximity to the probes to record the temperature. The sample is electrically insulated from the thermal block through a mica sheet. The entire unit is mounted at one end of the dip stick. The dip stick is dipped in a liquid nitrogen dewar for temperature dependent measurements. Temperature at the two ends of the sample  and the thermal voltage are measured using the Lakeshore 336 temperature controller and Keithley 2182A nanovoltmeter respectively. The entire system is LabVIEW automated.

\begin{figure}[H]
\centering
\includegraphics[width=12cm]{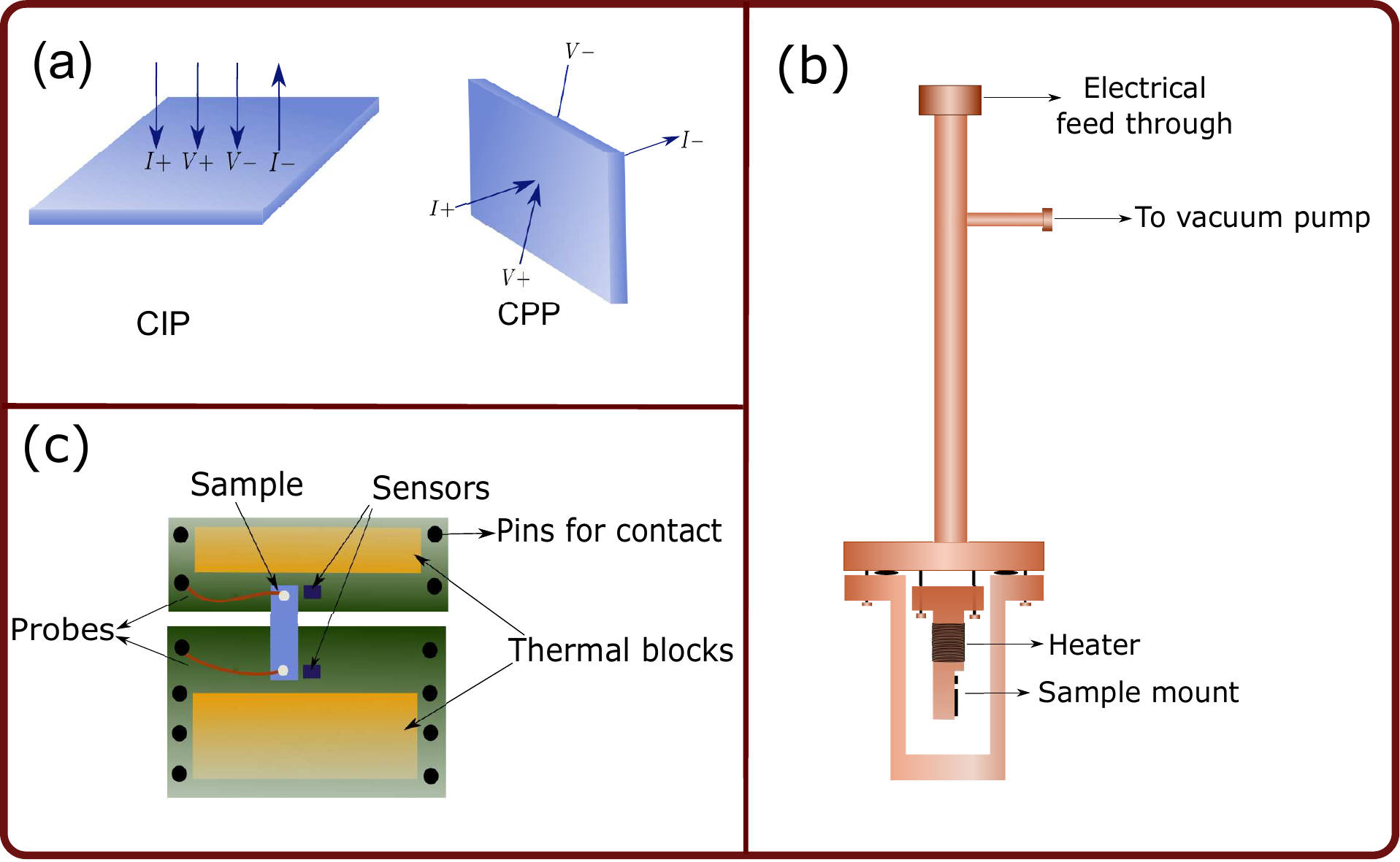}
\caption{(a) Schematic for CIP and CPP measurement configurations. (b) Schematic of  the Seebeck measurement dip-stick. (c) Front view of the sample mount. }
\end{figure}

\section{Results and Discussions}

Figure 2 (a) shows resistivity as a function of temperature in the range $9-290$ K in CIP mode in high vacuum of $10^{-4}$ mbar. Resistivity increases linearly in the range $9-28$ K, like in metals \cite{kittel} (Inset (i) Fig. 2 (a)). We observe a metal to semiconductor transition (Inset (ii) Fig. 2(a)) at $\simeq 31$ K. The resistivity behavior in the semiconducting phase can be explained by VRH model, according to which hopping between localized states is mediated by phonons. Above $32$ K resistivity falls with temperature as given by Eq. (1), where $d=2$, $\rho_0$ and $T_0$ are fitting parameters (Inset (ii) Fig. 2(a) shows raw data with VRH fitting of Eq. (1)). To show the quality of the fit, we also show  resistivity (in log scale) as a function of $T^{-1/3}$ in Fig. 2 (a). The value for $T_0$ obtained from this fit is  $\rm{3.7}\times 10^4$~K. This large value of $T_0$ is attributed to high degree of localization due to disorders. This value is comparable to the value reported in few layer 
MoSe$_2$ \cite{Roy,arindam}. This suggests that the localization length is small and the hence we observe VRH mechanism till room temperature in our experiment \cite{Roy,Arya,Shiraishi}.

We analyse the data for other possible transport mechanisms, details of which are discussed in supplementary information. An Arrhenius fit to the data gave impractically small magnitude of activation energy, $E_a$.  This suggests that the transport mechanism is different from thermionic excitations.

\begin{figure}[H]
\centering
\includegraphics[width=8cm]{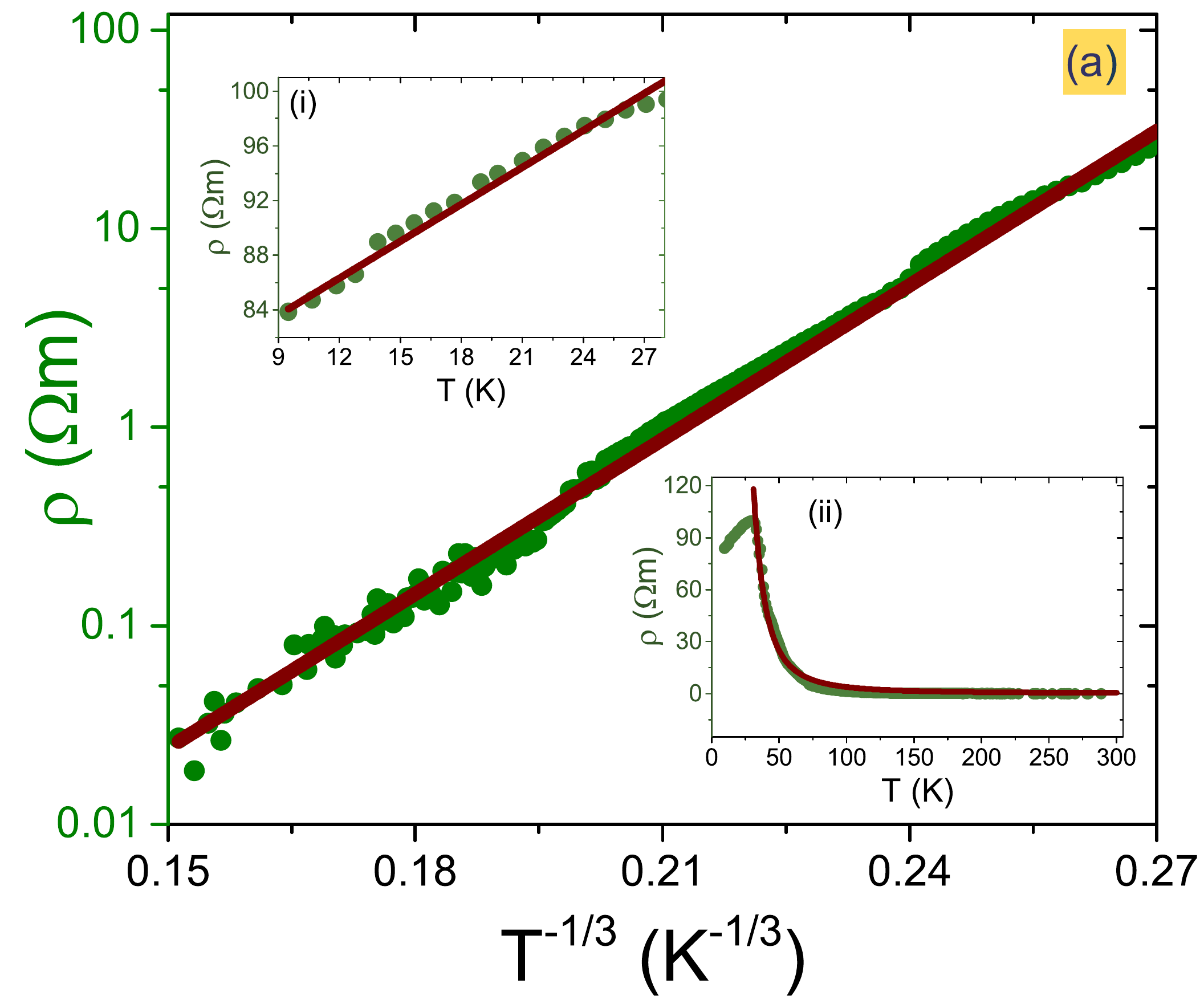}
\includegraphics[width=8cm]{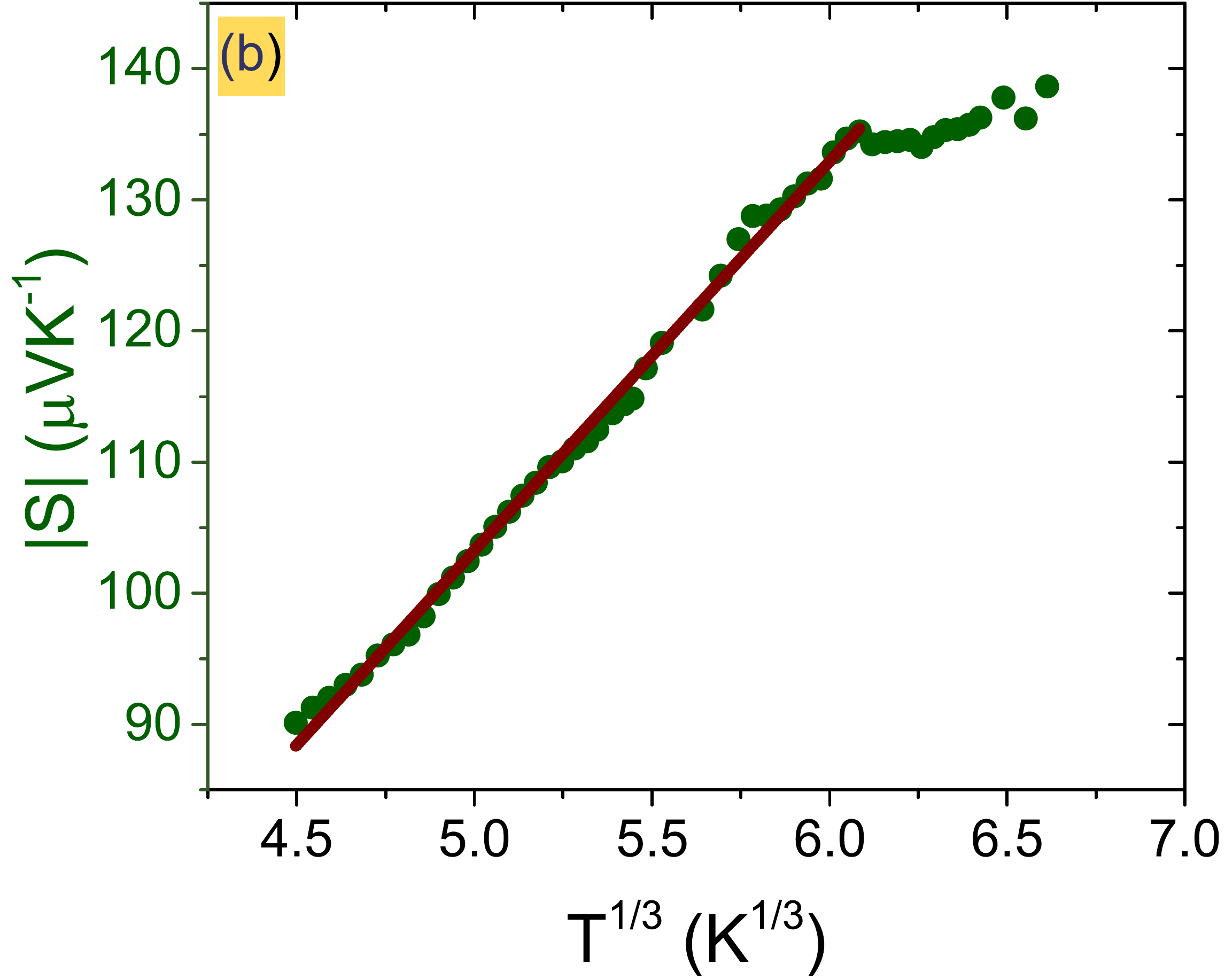}
\caption{(a) CIP mode Resistivity (in log scale) versus $T^{-\frac{1}{3}}$ in the range $32-290$ K. Inset (i)$\rho$ vs $T$ in the range $9-28$ K. Inset (ii) $\rho$ vs $T$ in the range $9-290$ K.  (b) In-plane Seebeck coefficient versus $T^{\frac{1}{3}}$ in the range $90-300$ K. Green circles show experimental data. Brown line represents the best fit curve. }
\end{figure}

 MoSe$_2$ has strong structural anisotropy (shown in further results for CPP measurements) and high degree of localization. As seen in fig. 2 (a), with decrease in temperature, divergence in resistivity is expected. However, we observe a down turn in resistivity, which varies linearly with temperature. This  semiconductor  to metal phase transition at $T \sim 31$ K observed in our experiment is highly intriguing which maybe attributed to disorder controlled divergence in resistivity as discussed by earlier works on phase transitions in disordered systems \cite{Lee2013,Dubi2007,PhysRevB.75.184530,Ying2016,Fujita2006}. In two dimensional quasi-1D systems of the MoSe$_2$ family, superconducting phase transition has been observed at lower temperatures \cite{Petrovic2016,Sarma,Singh, Bergk, Arut}. Petrovic et.al. \cite{Petrovic2016} show that for Na$_{2-\delta}$Mo$_6$Se$_6$, resistivity $\rho(T)$ as $T \rightarrow 0$ K, increases monotonically with degree of disorder. In our case disorder parameter $T_0$ is of the order of $10^4$, which corresponds to a finite $\rho$ at low temperature. Resistivity measurements at lower temperatures will reveal the emergent electronic phases of MoSe$_2$, whether it continues in metal like phase or undergo a superconducting phase transition assisted by Quantum phase slip. However the operational range of our experimental set-up limits the measurements to $9$ K. It must be noted that the value of resistivity observed is too large compared with that of normal metals ($\sim 10^{-8}$ $\Omega$m). This linear behavior maybe attributed to disorders, anisotropy and interactions between electrons \cite{crc} unlike electron-phonon scattering in normal metals.

Figure 2 (b) shows Seebeck coefficient versus temperature data in the range $90-300$~K. Temperature dependence of Seebeck coefficient is given by Eq. (3), where $d=2$ in our case. Seebeck coefficient in the VRH model has been explained in two ways. First, by the Mott picture where $S$ varies as stated above. In this case the $e-e$ interaction is not considered and resistivity is given by, Eq. (1). Second, the Efros picture considers $e-e$ interaction and predicts that below certain critical temperature the conductivity is given by $\sigma(T) = \sigma_0 \ exp \left(-\left(\frac{T_0}{T}\right)^\frac{1}{2}\right)$ \cite{chaikin}. This model predicts that the Seebeck coefficient is constant over the temperature range in which $e-e$ interactions are predominant. Figure~2~(b) shows Seebeck coefficient increase with temperature in accordance with Motts picture ($S \propto T^{1/3}$). Clearly we do not observe constant Seebeck coefficient nor the resistivity as per Efros' picture. The resistivity and the Seebeck both follow Motts model for electron and thermal transport. Hence we infer the dominant interaction in this temperature regime  ($T>31$ K) is electron-phonon interaction as explained by the Mott VRH model.

We find that the Seebeck coefficient increases as $T^{1/3}$ upto $225$ K. Beyond 225 K, the fit deviates due to effects of thermally generated charge carriers \cite{bentley}. In the whole temperature range the sign of Seebeck coefficient was negative. We find the Seebeck coefficient  to be $139 \ \mu$~V/K near to room temperature for bulk samples. This is significantly a large number for bulk sample in par with the ones which have been hitherto reported \cite{qdot,Imai}. Monolayer or few layer devices of MoSe$_2$ might have a much larger Seebeck coefficient than the bulk \cite{Lee,Delatorre}.

\begin{figure}[H]
\centering
\includegraphics[width=8cm]{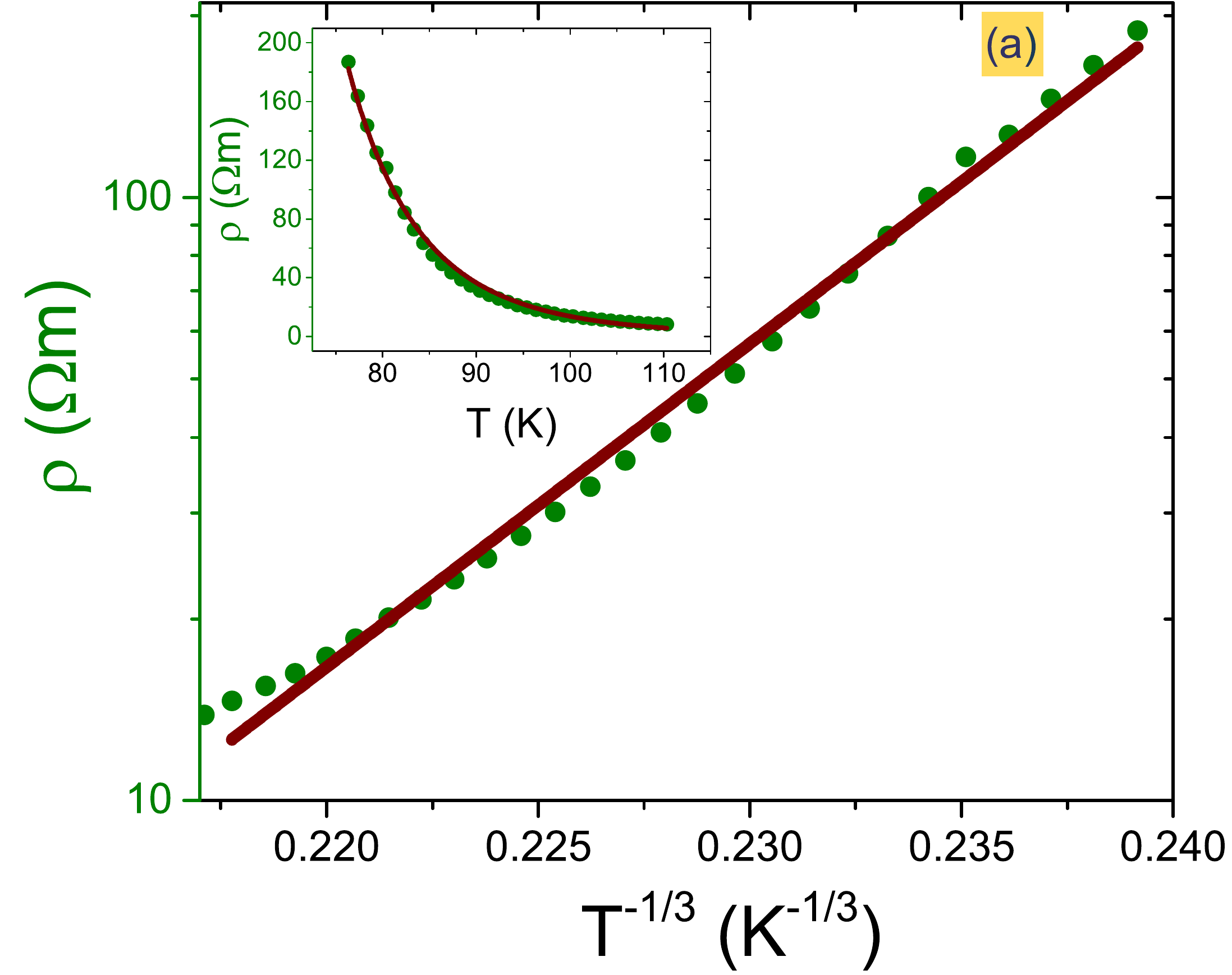}
\includegraphics[width=8cm]{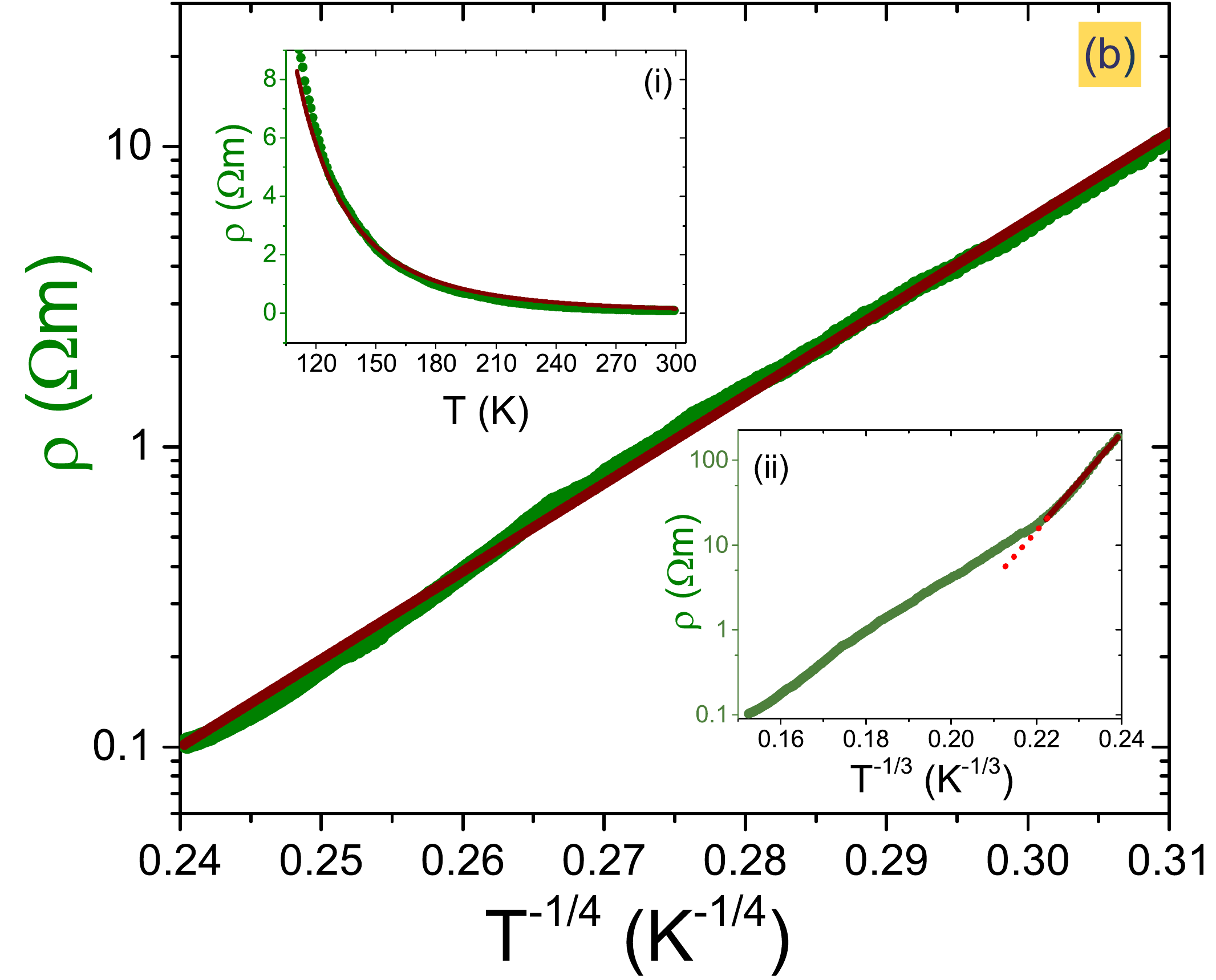}
\caption{CPP mode (a) Resistivity (in log scale) versus $T^{-1/3}$  in the range $76-106$ K. Inset shows $\rho$ versus $T$ in the same range. (b) Resistivity (in log scale) versus $T^{-1/4}$  in the range $107-300$ K. Inset (i) $\rho$ vs $T$ in the range $107-300$ K. Inset (ii) $\rho$ versus $T^{-1/3}$ in the range $76-300$ K,  $T^{-1/3}$ fit agrees only upto $106$ K. The red dashed line shows extrapolation of the $T^{-1/3}$ fit. For $T > 107$~K VRH goes as $T^{-1/4}$. Green solid circles are experimental data. The brown solid line represents the best fit curve. }
\end{figure}

Results of resistivity measurement for CPP mode in the range $76-300$ K are shown in Fig. 3. For temperatures below $76$ K, electrons freezing  limits the operational range of the setup. Due to strong structural anisotropy, magnitude of resistivity in CPP mode is at least three orders larger than that in CIP mode. The temperature dependence of resistivity in CPP configuration  varies from that in CIP mode. Figure 3 (a) shows resistivity data as a function of $T^{-1/3}$ in the range $76 -106$ K (Inset (i) shows $\rho$ $vs$ $T$).   The VRH model fitting as in Eq. (1) for CPP mode gives a  good fit for $d=2$ in this range; $T_0$ is of the order of $\sim 10^6$ K.  Hence we conclude that in this temperature range, transport is dominated by intralayer hopping.  Figure 3 (b) shows resistivity as a function of $T^{-1/4}$ in the range $107-300$ K in log scale (Inset shows $\rho$ $vs$ $T$). This  corresponds to resistivity with $d=3$ in the Mott VRH model ; $T_0$ is of the order of $\sim 10^7$ K. The crossover from $d=2$ to $d=3$ is explicitly shown in inset (ii) of Fig. 3 (b). We infer that the transport via hopping mechanism follows $3$D bulk like behavior at higher temperatures. In this temperature range interlayer hopping also contributes to the transport. The fitting parameter $T_0$ relates to  the degree of disorder in the system. We find large magnitude of $T_0$ in CPP than that in CIP which is as expected.

\section{Conclusions} 
To summarize, we have experimentally investigated the electron and thermal transport in MoSe$_2$ through temperature dependent resistivity and Seebeck measurements. The CIP resistivity shows  linear behavior with temperature upto $31$ K.  For $T > 31$ K,  we have performed fitting according to the VRH model and observed that the transport in the intrinsic semiconducting phase is governed by localized states. Large magnitude of $T_0$, in the VRH model fit indicates high degree of localization. Temperature dependent resistivity measurements for $T < 9$ K will answer the curiosity on the phases that might occur at low temperatures. Resistivity in CPP mode is much larger ($\sim$ 3 orders) than the CIP mode, which clearly indicates strong structural anisotropy. The transport occurs in two dimensional regime upto $106$ K and exhibits bulk behavior above $107$ K.  In-plane Seebeck measurements also show VRH transport mechanism. For a bulk material the room temperature Seebeck coefficient is significantly large, and in par with materials like MoS$_2$.  Seebeck coefficient is expected to have much larger magnitude for thinner or few layer samples. Hence this material is a highly potential candidate for engineering heterostructure devices based on the PGEC concept.

\section{Supplementary Material}

Supplementary information contains discussion for weak localization and Arrhenius excitation for the experimental data in Fig. 2 (a) and XRD spectrum of MoSe$_2$ crystals.

\section{Acknowledgements} DS thanks Department of Science and Technology (DST), Govt. of India for PhD fellowship through DST-INSPIRE scheme (DST/INSPIRE Fellowship/2013/742). RSP thanks DST, Govt. of India for financial support (No. SR/NM/MS-1002/2010) through Nanomission program. We thank Abhiram Soori, Diptiman Sen and Radhika Vathsan  for illuminating discussions.

\bibliography{references}

\end{document}